\documentclass[epsfig,12pt] {article}
\usepackage{graphics}
\usepackage{epsfig}
\usepackage{amssymb}
\usepackage{amsmath}
\usepackage{bm}
\usepackage{epsfig}

\begin{document}

\baselineskip .7cm

\author{ Navin Khaneja \thanks{To whom correspondence may be addressed. Email:navinkhaneja@gmail.com} \thanks{Department of Electrical Engineering, IIT Bombay, Powai - 400076, India.}}

\vskip 4em

\title{\bf Double swept band selective excitation}

\maketitle

\vskip 3cm

\begin{center} {\bf Abstract} \end{center}
The paper describes the design of band selective excitation and rotation pulses in high resolution NMR by method of double sweep.  We first show the design of a pulse sequence that produces band selective excitation to the equator of Bloch sphere with phase linearly dispersed as frequency. We show how this linear dispersion can then be refocused by nesting free evolution between two adiabatic inversions (sweeps). We then show how this construction can be generalized to give a band selective $x$ rotation over desired frequency band. Experimental excitation profiles for the residual HDO signal in a sample of $99.5\%$ D$_2$O are obtained as a function of resonance offset.

\vskip 3em

\section{Introduction}

Frequency-selective pulses have widespread use in magnetic resonance and significant effort has been devoted towards their design \cite{barrett}-\cite{glaser}.
Several experiments in high-resolution NMR and magnetic resonance
imaging require radiofrequency pulses which excite NMR response over a prescribed frequency range with negligible effects elsewhere.
Such band-selective pulses are particularly valuable when the excitation is uniform over desired bandwidth and of constant phase.

In this paper, we propose a new approach for design of uniform phase, band selective excitation and rotation pulses. In this approach, using Fourier series, a pulse sequence that produces band selective excitation to equator of Bloch sphere with phase linearly dispersed as frequency is designed. This linear dispersion is then refocused by nesting free evolution between two adiabatic inversions (sweeps). This construction is generalized to give a band selective $x$-rotation over desired bandwidth. We assume uncoupled spin $\frac{1}{2}$ and neglect relaxation.

The paper is organized as follows. In section 2, we present the theory behind double swept {\bf ba}nd{\bf s}elective {\bf e}xcitation, we call {\bf BASE}. In section 3, we present simulation results and experimental data for band selective
excitation and rotation pulses designed using double sweep technique. Finally, we conclude in section 4, with discussion and outlook.

\section{Theory}

We consider the problem of band selective excitation. Consider the evolution of spinor(We use $I_\alpha$ to denote the Pauli matrix such that $\alpha \in \{x, y, z \}$ ) of a spin $\frac{1}{2}$ in a rotating frame, rotating around $z$ axis at Larmor frequency.

\begin{equation}
\label{eq:basiceq}
\frac{d |\psi \rangle }{dt} = -i ( \omega I_z + A(t) \cos \theta(t) I_x + A(t) \sin \theta(t) I_y )  |\psi \rangle,
\end{equation}where $A(t)$ and $\theta(t)$ are amplitude and phase of rf-pulse and we normalize the chemical shift in the range $\omega \in [-1, 1]$. Our goal is to design excitation over the bandwidth $[-B, B]$ where $B < 1$. Let $u(t) = A(t) \exp(-j \theta)$.

\begin{equation}
\frac{d |\psi \rangle }{dt} = -\frac{i}{2} \left [ \begin{array}{cc} \omega & u(t)\\ u^\ast(t) & -\omega \end{array} \right ] |\psi \rangle.
\end{equation}

Going into interaction frame of chemical shift, we can write the evolution as

\begin{equation}
\label{eq:solution}
|\psi(T) \rangle = \left [ \begin{array}{cc} \exp(\frac{-i \omega T}{2})  & 0 \\ 0 & \exp(\frac{i \omega T}{2}) \end{array} \right ] \exp (\oint_0^T \frac{-i}{2} \left [ \begin{array}{cc} 0 & u(t) \exp(i \omega t) \\ u^\ast(t) \exp(-i \omega t) & 0 \end{array} \right ]) |\psi(0) \rangle.
\end{equation}

We write,

\begin{equation}
\int_0^T  u(t) \exp(i \omega t) = \exp( \frac{i \omega T}{2}) \int_0^T  u(t) \exp(i \omega (t - \frac{T}{2})).
\end{equation}

We design $u(t)$ such that for all $\omega \in [-B, B]$ we have

\begin{equation}
 \int_0^T  u(t) \exp(i \omega (t - \frac{T}{2})) \sim \frac{\pi}{2}.
\end{equation} and zero elsewhere.

Divide $[0, T]$ in intervals of step, $\Delta t$, over which $u(t)$ is constant. Call them $\{u_{-M}, \dots, u_{-k}, \dots, u_0 \}$ over
$[0, \frac{T}{2}]$ and  $\{u_{0}, \dots, u_{k}, \dots, u_M \}$ over $[\frac{T}{2}, T]$.

\begin{equation}
\label{eq:fourierapprox}
 \int_0^T  u(t) \exp(i \omega (t - \frac{T}{2})) \sim (u_0 + \sum_{k = -M}^{M} u_k \exp(-i \omega k \Delta t))\Delta t, 
\end{equation}

where write $\Delta t = \frac{\pi}{N}$ and choose $u_k$ real with $u_k = u_{-k}$. Then we get 

\begin{equation}
\label{eq:fourierseries}
 \int_0^T  u(t) \exp(i \omega (t - \frac{T}{2})) \sim 2 \sum_{k = 0}^{M} u_k \cos (\omega k \Delta t) \Delta t =  2 \sum_{k = 0}^{M} u_k \cos (k x) \Delta t,
\end{equation} where for $x \in [ -\frac{\pi B}{N}, \frac{\pi B}{N} ]$, we have $2 \sum_{k = 0}^{M} u_k \cos (k x) \Delta t \sim \frac{\pi}{2}$ and $0$ for $x$ outside this range. This is a Fourier series, and we get the Fourier coefficients as, 

\begin{equation}
\label{eq:fouriercoeff}
u_0 = \frac{B}{4}\ ; \ \ u_k = \frac{\sin(\frac{k B \pi}{N})}{\frac{2 k \pi}{N}}.
\end{equation}

Approximating, for $\omega \in [-B, B]$,

\begin{eqnarray}
\label{eq:frame1}
\nonumber \exp (\oint_0^T \left [ \begin{array}{cc} 0 & \frac{-i u(t) \exp(i \omega t)}{2} \\ \frac{-i u^\ast(t) \exp(-i \omega t)}{2} & 0 \end{array} \right ]) &\sim&  \exp(  \frac{-i}{2} \left [ \begin{array}{cc} 0 & \int_0^T u(t) \exp(i \omega t) \\ \int_0^T u^\ast(t) \exp(-i \omega t) & 0 \end{array} \right ])\\ \nonumber &\sim& \exp(  \frac{-i}{2} \left [ \begin{array}{cc} 0 & \exp( \frac{i \omega T}{2}) \frac{\pi}{2} \\  \exp( \frac{-i \omega T}{2}) \frac{\pi}{2} & 0 \end{array} \right ]) \\ &=& \exp(-i \frac{\pi}{2} (\cos \frac{\omega t}{2} I_x - \frac{\sin \omega t}{2} I_y)) \\ &=&   \frac{1}{\sqrt{2}} \left [ \begin{array}{cc} 1 & -i \exp( \frac{i \omega T}{2})\\  -i \exp( \frac{-i \omega T}{2}) & 1 \end{array} \right ].
\end{eqnarray}

Starting from the initial state $|\psi(0) \rangle = \left [ \begin{array}{c} 1 \\ 0 \end{array} \right ]$, from Eq. \ref{eq:solution}, 
for $\omega \in [-B, B]$, we have

\begin{equation}
|\psi(T) \rangle = \frac{1}{\sqrt 2} \left [ \begin{array}{c} \exp( \frac{-i \omega T}{2}) \\ -i  \end{array} \right ].
\end{equation} There is no excitation outside desired band.

This state is dephased on the Bloch sphere equator. We show how using a double adiabatic sweep, we can refocus the phase. Let $\Theta(\omega)$ be the rotation for an adiabatic inversion of a spin. We can use Euler angle decomposition to write, 

\begin{equation}
\Theta(\omega) = \exp(-i \alpha(\omega) I_z) \exp(-i \pi I_x) \exp(-i \beta(\omega) I_z).
\end{equation} The center rotation should be $\pi$ for $\Theta(\omega)$ to do inversion of $I_z \rightarrow -I_z$.

We can use this to refocus the forward free evolution. Observe

\begin{equation}
\Delta (\omega, \frac{T}{2}) = \exp(\frac{i \omega T}{2} I_z) = \Theta(\omega) \exp(\frac{-i \omega T}{2} I_z) \Theta(\omega).
\end{equation}

Then for $\omega \in [-B, B]$,

\begin{equation}
\label{eq:firstexcitation}
\Theta(\omega) \ \exp(\frac{-i \omega T}{2} I_z) \ \Theta(\omega) \ \frac{1}{\sqrt 2} \left [ \begin{array}{c} \exp( \frac{-i \omega T}{2}) \\ -i  \end{array} \right ] = \frac{\exp( \frac{-i \omega T}{4})}{\sqrt 2} \left [ \begin{array}{c} 1 \\ -i  \end{array} \right ],
\end{equation} which is a band selective excitation.

\begin{figure}
\centering
\includegraphics[scale=.6]{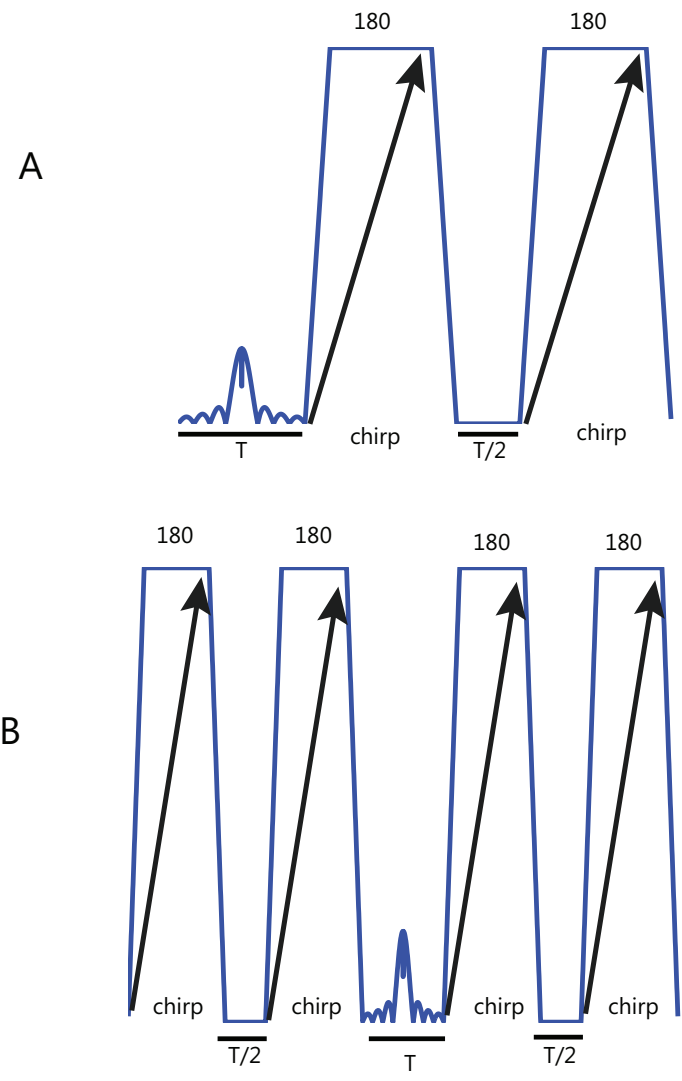}
\caption{Figure A. shows the {\bf BASE} pulse sequence (amplitude) with a double sweep that performs band selective excitation as in Eq. (\ref{eq:firstexcitation}) for $B = \frac{1}{5}$.
Figure B. shows the {\bf BASE} pulse sequence with two double sweeps that performs band selective rotation as in Eq. (\ref{eq:firstrotation}) for $B= \frac{1}{5}$.}\label{fig:pulsesequence}
\end{figure}

The pulse sequence consists of a sequence of x-phase pulses, which produce the evolution

\begin{equation}
U(\omega, \theta) = \left [ \begin{array}{cc} \exp(\frac{-i \omega T}{2})  & 0 \\ 0 & \exp(\frac{i \omega T}{2}) \end{array} \right ]\exp(  \frac{-i}{2} \left [ \begin{array}{cc} 0 & \exp( \frac{i \omega T}{2}) \theta \\  \exp( \frac{-i \omega T}{2}) \theta & 0 \end{array} \right ]),
\end{equation} where $\theta = \frac{\pi}{2}$, as described above. This required a 
peak amplitude of $u(t) \sim \frac{B}{2}$. This is followed by a double sweep rotation $\Delta (\omega, \frac{T}{2})$. Fig. \ref{fig:pulsesequence}A shows the pulse sequence for $B= \frac{1}{5}$. The sweep(chirp) is done with a peak amplitude of $\frac{1}{2}$, $T = 40 \pi$.

We talked about band selective excitations. Now we discuss band selective $\frac{\pi}{2}$ rotations. This is simply obtained from above by an initial double sweep.
Thus

\begin{equation}
\label{eq:firstrotation}
U_1 = \Delta (\omega, \frac{T}{2}) \ U(\omega, \frac{\pi}{2}) \ \Delta (\omega, \frac{T}{2}),
\end{equation} is a $\frac{\pi}{2}$ rotation around $x$ axis.  Fig. \ref{fig:pulsesequence}B shows the band selective rotation pulse sequence for $B=.2$. The chirp is done with a peak amplitude of $\frac{1}{2}$, $T = 40 \pi$.

 \section{Simulations}

\begin{figure}[htb]
\centering
\begin{tabular}{cc}
\includegraphics[scale=.25]{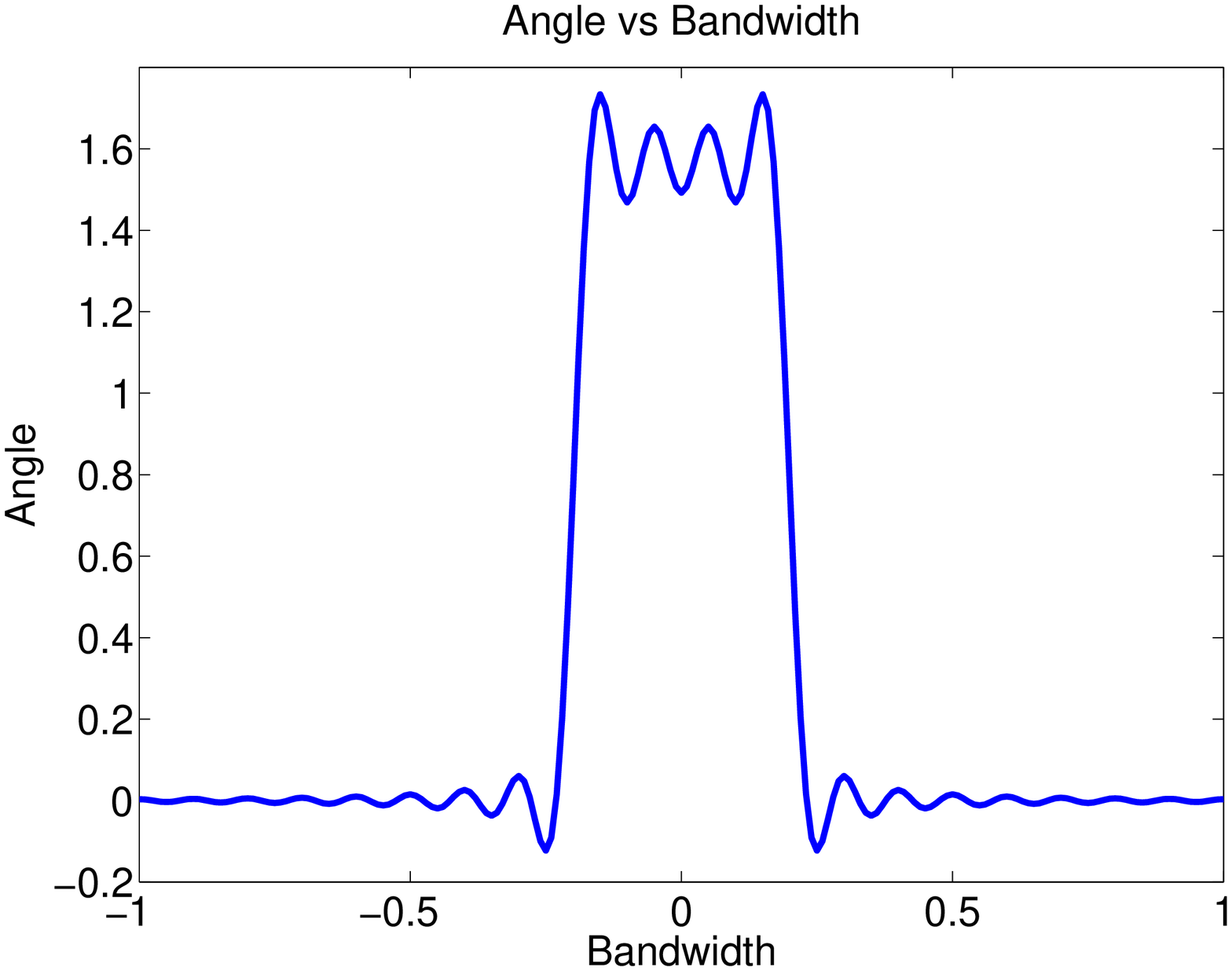} & \includegraphics[scale=.25]{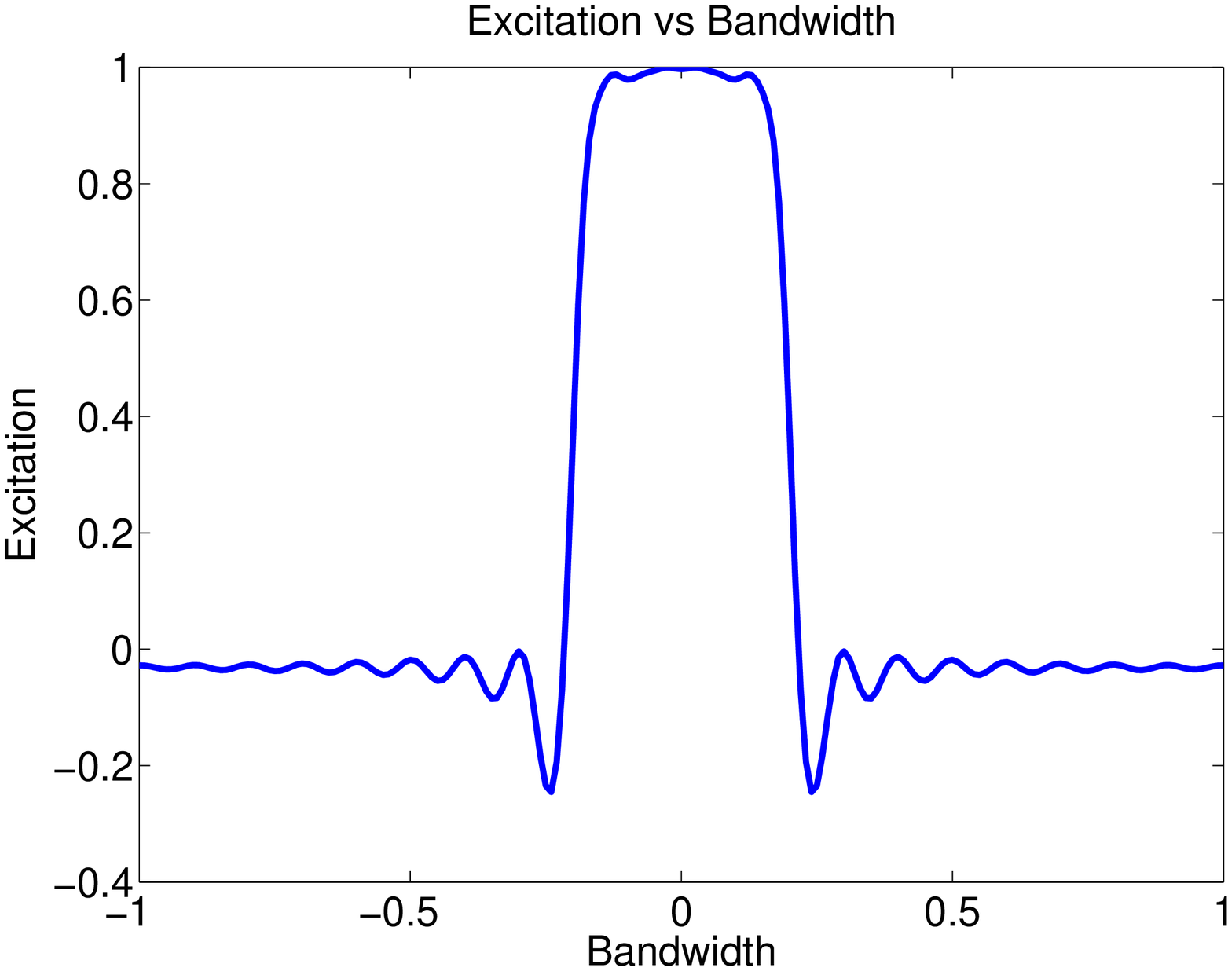}
\end{tabular}
\caption{Left panel shows value of the Eq. (\ref{eq:fourierseries}) as a function of bandwidth when we choose $T = 40 \pi$ and $\Delta t = \frac{\pi}{10}$, $B = \frac{1}{5}$. 
The right panel shows the excitation profile i.e., the $-y$ coordinate of the Bloch vector after application of the pulse in Eq. (\ref{eq:firstexcitation}), with $u_k$ as in Eq. (\ref{eq:fouriercoeff}) and we assume that adiabatic inversion is ideal.}\label{fig:fourierangle}
\end{figure}

\begin{figure}[!htb]
\centering
\includegraphics[scale=.7]{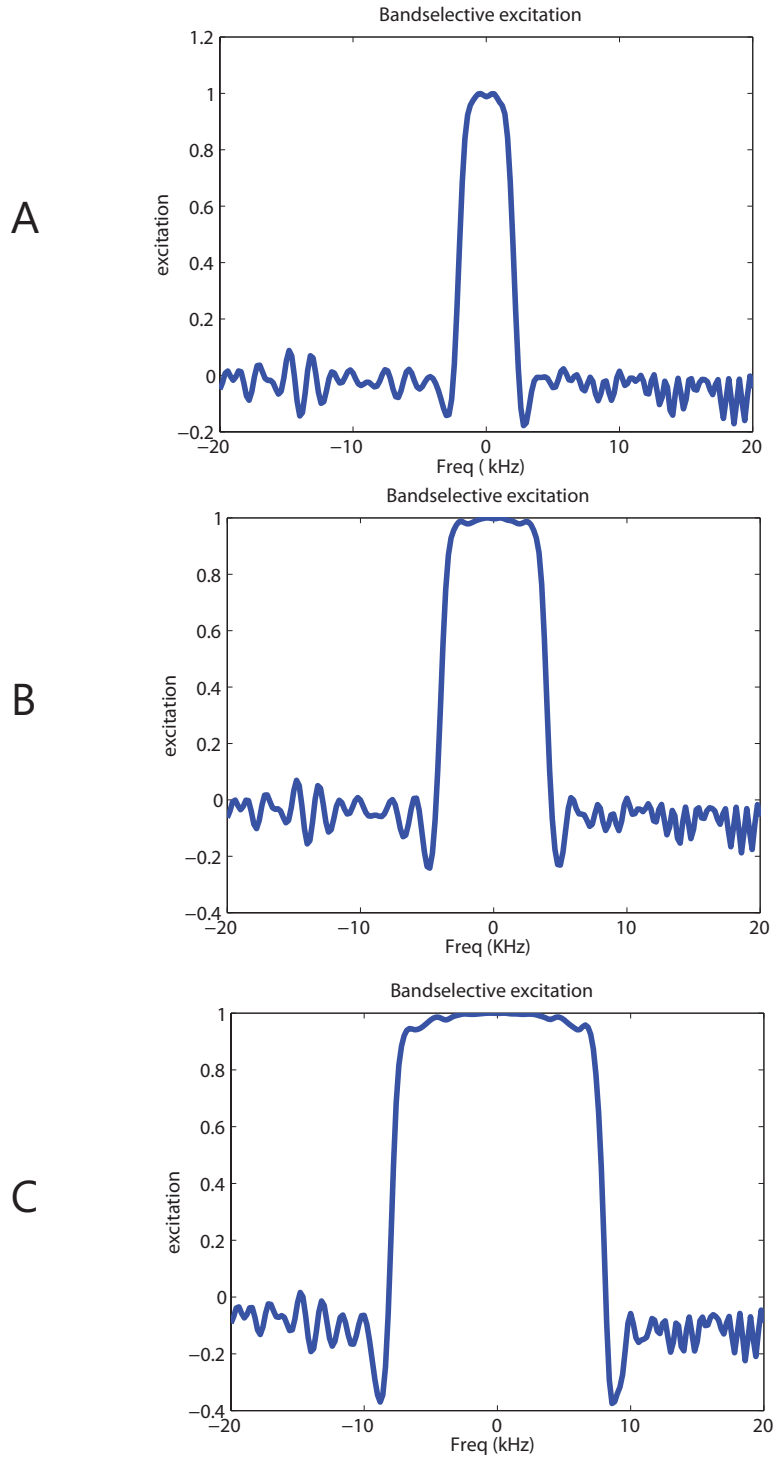} 
\caption{Fig. A, B, C shows the excitation profile (the $-y$ coordinate of Bloch vector) for the {\bf BASE} pulse in Eq. (\ref{eq:firstexcitation}) with $B = [-2, 2]$ kHz, $B = [-4, 4]$ kHz and $B = [-8, 8]$ kHz, respectively. The peak amplitude is $A = 10$ kHz. Time of the pulse is $3.89$ ms. $T= 1$ ms in Fig. \ref{fig:pulsesequence}A. }\label{fig:excitation}
\end{figure}

\begin{figure}[!htb]
\centering
\includegraphics[scale=.7]{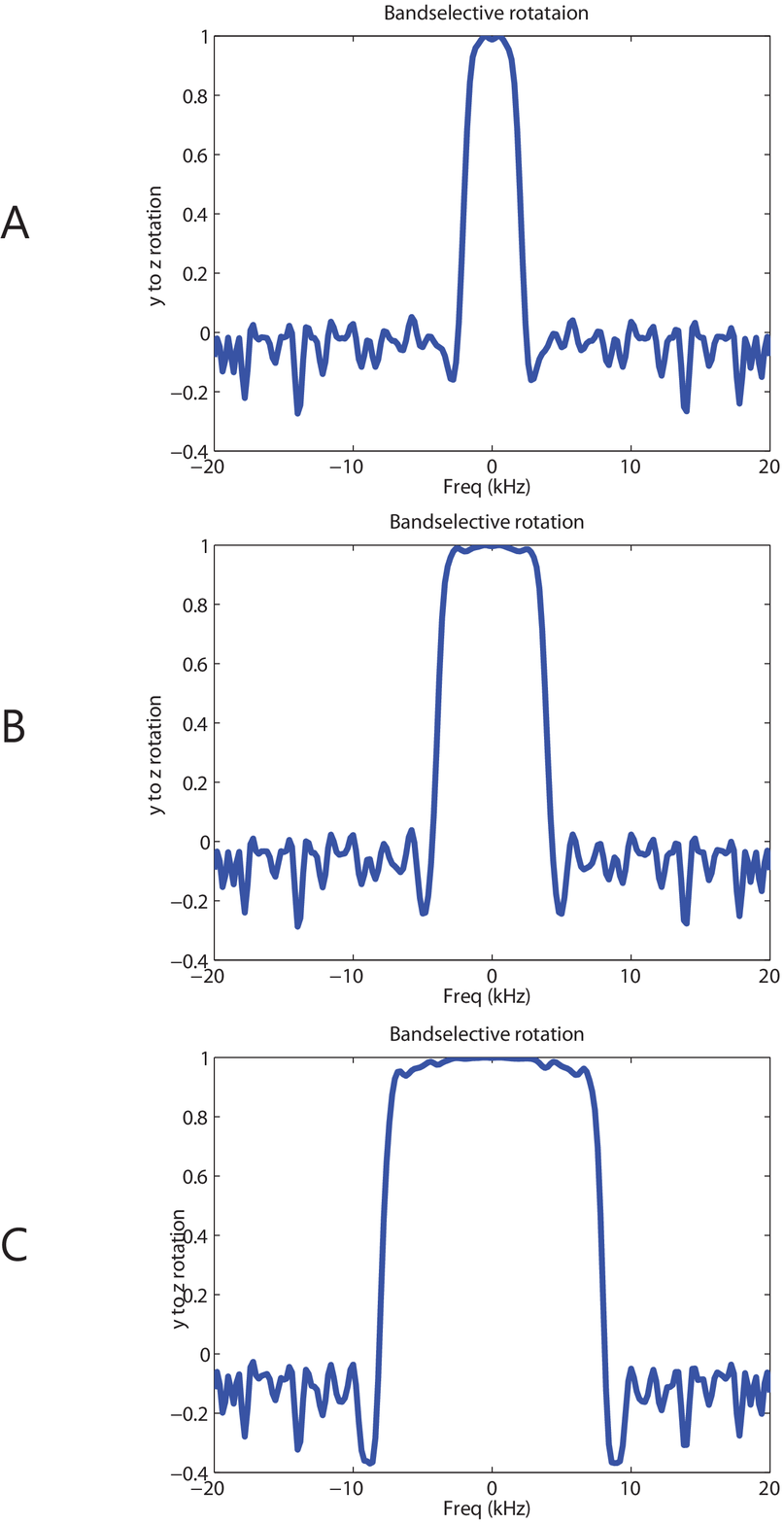}
\caption{Fig. A, B, C shows the the $y$ to $z$ rotation profile (the $z$ coordinate of Bloch vector) for the band selective $x$ rotation pulse in Eq. (\ref{eq:firstrotation}) with $B = [-2, 2]$ kHz, $B = [-4, 4]$ kHz and $B = [-8, 8]$ kHz, respectively. The peak amplitude is $A = 10$ kHz. Time of the pulse is $6.77$ ms. $T= 1$ ms in Fig. \ref{fig:pulsesequence}B.}\label{fig:rotation}
\end{figure}

We normalize $\omega$ in Eq. (\ref{eq:basiceq}), to take values in the range $[-1, 1]$. We choose time $\frac{T}{2} = M \pi$, where we choose $M = 20$ and $N = 10$ in 
$\Delta t = \frac{\pi}{N}$ in Eq. (\ref{eq:fourierapprox}). Choosing $\theta = \frac{\pi}{2}$ and coefficients $u_k$ as in Eq. (\ref{eq:fouriercoeff}), we get the value of the
Eq. (\ref{eq:fourierseries}) as a function of bandwidth as shown in left panel of Fig. \ref{fig:fourierangle} for $B = .2$. This is a decent approximation to $\frac{\pi}{2}$ over the desired bandwidth. The right panel of Fig. \ref{fig:fourierangle}, shows the excitation profile i.e., the $-y$ coordinate of the Bloch vector after application of the pulse in Eq. (\ref{eq:firstexcitation}), where we assume that adiabatic inversion is ideal. The peak rf-amplitude $A \sim \frac{B}{2}$ for $B = .2$.

Next, we implement the nonideal adiabatic sweep with a chirp pulse, by sweeping from $[-1.5, 1.5]$ in $150$ units of time. This gives a sweep rate $\frac{1}{50} \ll A^2$, where $A = \frac{1}{2}$. The chirp pulse is a depicted in Fig.  \ref{fig:pulsesequence}. The chirp operates at its peak amplitude over sweep from $[-1, 1]$. The resulting excitation profile of Eq. (\ref{eq:firstexcitation}) is shown in  Fig. \ref{fig:excitation} A, where we show the $-y$ coordinate of the Bloch vector.
After scaling,  $\omega \in [-20, 20]$ kHz, $B = 2$ kHz and $A = 10$ kHz, this pulse takes $3.89$ ms.  In Fig. \ref{fig:excitation} B, and \ref{fig:excitation} C, we have $B = 4$ kHz and $B = 8$ kHz respectively. The pulse time is same $3.89$ ms.
 $T= 1$ ms in Fig. \ref{fig:pulsesequence}A.

Next, we simulate the band selective $x$ rotation as in Eq. (\ref{eq:firstrotation}). This requires to perform double sweep twice as in  Eq. (\ref{eq:firstrotation}). Adiabatic sweep is performed as before. The resulting excitation profile of Eq. (\ref{eq:firstrotation}) is shown in  Fig. \ref{fig:rotation} A,B and C, where we show the $z$ coordinate of the Bloch vector starting from initial $y = 1$ for $B = [-2, 2]$ kHz, $B = [-4, 4]$ kHz and $B = [-8, 8]$ kHz respectively.
This pulse takes $6.77$ ms in each case.

\subsection{Experimental}

\begin{figure}[!htb]
\centering
\includegraphics[scale=.5]{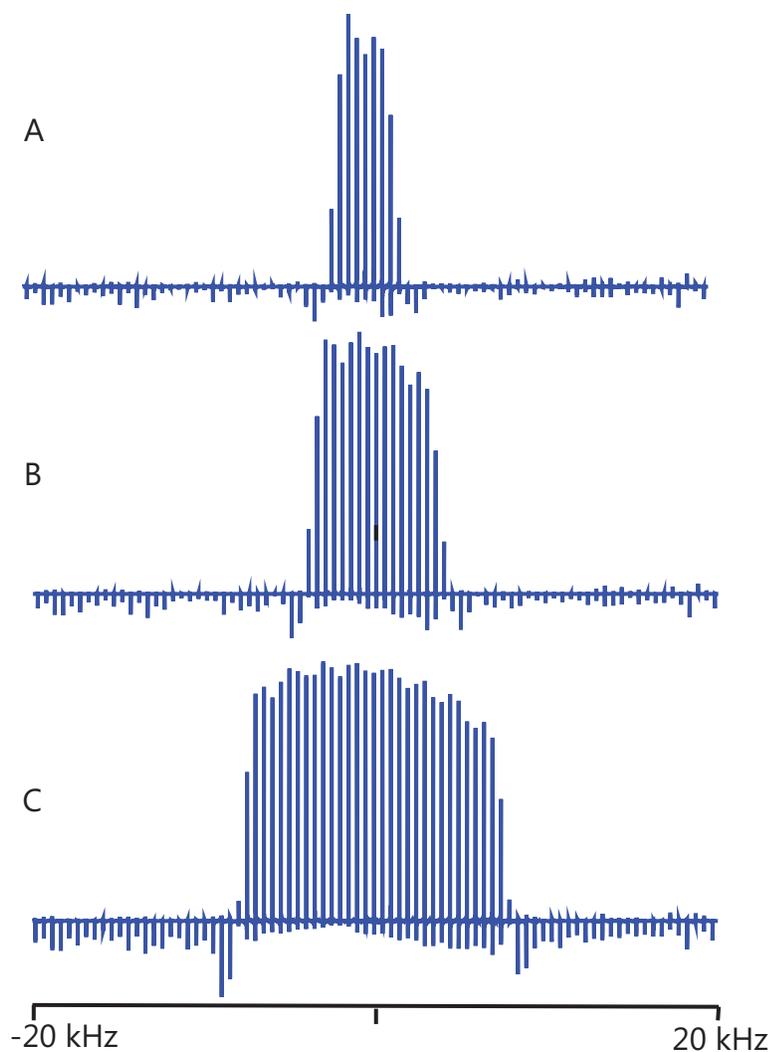}
\caption{Fig. A, B, C show the experimental excitation profile of {\bf BASE} sequences in Fig. \ref{fig:excitation}A,B and C, respectively, 
 with $B = [-2, 2]$ kHz, $B = [-4, 4]$ kHz and $B = [-8, 8]$ kHz, respectively, in a sample of $99.5\%$ D$_2$O. The offset is varied over the range as shown and the peak rf power of all pulses is 10 kHz. The duration of the pulses is $3.89$ ms. }\label{fig:dsweep}
\end{figure}

All experiments were performed on a 750 MHz (proton
frequency) NMR spectrometer at 298 K.
Fig. \ref{fig:dsweep} shows the experimental excitation profiles for the residual HDO signal in a sample
of $99.5\%$ D$_2$O displayed as a function of resonance offset. Fig.  \ref{fig:dsweep}A, B, C shows the excitation profile of {\bf BASE} sequence in   Fig. \ref{fig:excitation} A, B, C respectively.  The frequecy band of interest is $[-2, 2]$ kHz, $[-4, 4]$ kHz and $[-8, 8]$ kHz respectively. In each case, the peak amplitude of 
the rf-field is 10 kHz and duration of the pulse is 3.89 ms. The pulse sequence uses one double sweep. $T= 1$ ms in Fig. \ref{fig:pulsesequence}A. The offset is varied over a range of [-20, 20] kHz with on-resonance at 3.53 kHz (4.71 ppm). 

\section{Conclusion}

In this paper we showed design of band selective excitation and rotation pulses ({\bf BASE}). We first showed how by use of Fourier series, we can design a pulse that does band selective excitation to the equator of Bloch sphere. The phase of excitation is linearly dispersed as function of offset, which is refocused by nesting free evolution between adiabatic inversion pulses. We then extended the method to produce band selective rotations. The pulse duration of the pulse sequences is largely limited by time of adiabatic sweeps. This increases, if we have larger working bandwidth. However, we can invert only the band of interest. Thereby, we may be able to reduce the time of the proposed pulse sequences. The principle merit of the proposed pulse sequences is the analytical tractability and conceptual simplicity of the design.

\section{Acknowledgement}
The author would like to thank the HFNMR lab facility at IIT Bombay, funded by RIFC, IRCC, where the data was collected.

\end{document}